\documentstyle[amssymb]{article}

\begin{document}

\begin{center}

{\bf \Large Non-Laplacian growth, algebraic domains and finite reflection groups}\\

\vspace{5mm}

{\Large Igor Loutsenko, Oksana Yermolayeva}

\vspace{5mm}

Mathematical Institute, University of Oxford, 24-29 st. Gilles',
Oxford, OX1 3LB, UK, e-mail: loutsenk@maths.ox.ac.uk

 \vspace{5mm}

SISSA, via Beirut 2-4, Trieste, 34014, Italy, e-mail:
yermola@sissa.it

Abstract

\begin{quote}

Dynamics of planar domains with moving boundaries driven by the
gradient of a scalar field that satisfies an elliptic PDE is
studied. We consider the question: For which kind of PDEs the
domains are algebraic, provided the field has singularities at a
fixed point inside the domain? The construction reveals a direct
connection with the theory of the Calogero-Moser systems related to
finite reflection groups and their integrable deformations.

\end{quote}

\end{center}

\begin{section}{Introduction, non-homogeneous porous medium flows}

Laplacian growth is a process that governs the dynamics of the
boundary $\partial\Omega=\partial\Omega(t)$ in the plane separating
two disjoint, open regions ($\Omega$ and ${\mathbb
C}\backslash\bar\Omega$) in which harmonic (scalar) fields are
defined. These may be interpreted as the pressure fields for
incompressible fluids (porous-medium flows, Hele-Shaw flows, etc.
See e.g. \cite{VE} and references therein). In the recent literature
there have appeared several new formulations of the Laplacian growth
related with the theory of integrable systems and random matrices,
quantum Hall effect and Dirichlet boundary problem in two-dimensions
(see e.g. \cite{HLY}, \cite{KMZ}, \cite{Z} and references therein).

In the present paper, we consider a new connection with the theory
of quantum integrable systems. To be more precise, we study an
integrable generalization of the Laplacian growth, when the boundary
is driven by a field satisfying an elliptic PDE, that is not
generally reduced to a Beltrami-Laplace equation ("Non-Laplacian"
growth). To be specific, we use the porous medium fluid dynamics
interpretation.

We find variable-coefficient elliptic PDEs for which the boundary
dynamics can be described explicitly and the moving fluid occupies
evolving algebraic domains (see below). These, turn out, to be PDEs
of the Calogero-Moser type, related to finite reflection (Coxeter)
groups as well as their integrable deformations (see Conclusion),
that possibly complete the list of all second-order PDEs connected
with the algebraic domains (our main conjecture).

In this section we set the problem in terms of the porous medium
fluid dynamics. Formulation of the problem in terms of quadrature
domains is given in the next section.

Consider a flow of a liquid in a thin non-planar layer of
non-homogeneous porous medium. The layer can be viewed as a
2-dimensional surface embedded in the three-dimensional Euclidean
space. We let the layer curvature, permeability, porosity and
thickness depend on the surface spatial coordinates $x,y$. We can
choose $x,y$ such that locally
$$
dl^2=G(x,y)(dx^2+dy^2)
$$
where $dl$ is the surface length element. The surface area element
is $d\Sigma=G dxdy$, and the volume of the liquid that can be
absorbed in the range $x+dx,y+dy$ equals
$$
dV=\eta h d\Sigma=\eta hGdxdy,
$$
where $\eta=\eta(x,y), h=h(x,y)$ are the medium porosity and the
layer thickness, respectively.

In the porous medium, the flow velocity $v=(dx/dt,dy/dt)$ is
proportional to the gradient $\nabla=(\partial/\partial x,
\partial/\partial y)$ of the pressure $P$
$$
v=-\frac{\kappa}{\sqrt{G}}\nabla P,
$$
where $\kappa=\kappa(x,y)$ is the medium permeability.

It is seen from the above that only the two combinations of variable
coefficients, namely
$$
\eta h G, \quad \frac{\kappa}{\sqrt{G}}
$$
enter the flow equation of motion, and it is convenient to absorb
$h$ and $G$ into definitions of the other coefficients. Therefore,
without loss of generality, we can consider the flow in the plane
parametrized by the complex coordinates $z=x+iy, \bar z=x-iy$,
choosing $\eta$ and $\kappa$ to depend on $z, \bar z$, while setting
remaining coefficients to unity. The liquid volume conservation
leads to the continuity equation
\begin{equation}
(\nabla\cdot\eta v)=0 \label{continuity},
\end{equation}
while the dynamical law of motion rewrites as
\begin{equation}
v=-\kappa\nabla P \label{darcy}.
\end{equation}
We consider a situation where the liquid occupies a bounded,
simply-connected open region $\Omega$ of the plane, whose time
evolution $\Omega=\Omega(t)$ is induced by the flow.

At fixed time $t$, the pressure is constant along the boundary
\begin{equation}
P(\partial\Omega(t))=P_0(t). \label{boundary}
\end{equation}
Note, that the in the case of the simply-connected domains
considered here, the dynamics is independent of $P_0(t)$, and for
convenience the latter can be set to zero.

The normal velocity of the boundary $v_n$ and that of the flow
coincide at $\partial\Omega$
\begin{equation}
v_n=n\cdot v\quad {\rm if} \quad z\in
\partial\Omega \label{kinematik}.
\end{equation}
The flow is singularity driven. For instance
\begin{equation}
P\to \frac{-q(t)}{2\kappa(z_1,\bar z_1)\eta(z_1,\bar
z_1)}\log|z-z_1|+\sum_{j=1}^
k\left(\frac{\mu_j(t)}{(z-z_1)^j}+\frac{\bar\mu_j(t)}{(\bar z-\bar
z_1)^j}\right), \quad {\rm as} \quad z \to z_1, \label{sources}
\end{equation}
when a multipole source of order $k+1$ is located at $z=z_1$.
Equations (\ref{continuity}) - (\ref{sources}) constitute a free
boundary problem where the evolution of the boundary $\partial
\Omega(t)$ is completely determined by the initial condition
$\partial \Omega(0)$ and strengths $q(t)=\bar q(t), \mu_j(t), \bar
\mu_j(t), j=1,..,k$ as well as position $z_1$ of the sources.

\end{section}

\begin{section}{Conservation Laws and Quadrature Domains}

From (\ref{continuity}), (\ref{darcy}), (\ref{sources}) it follows
that the pressure satisfies the elliptic PDE
\begin{equation}
\nabla\kappa\eta\nabla P=-\pi\hat q[\delta(x-x_1)\delta(y-y_1)]
\label{pressure}.
\end{equation}
where $\hat q=\hat q(t)$ is the differential operator of order $k$
\begin{equation}
\hat q=q(t)+\sum_{j=1}^k (-1)^j\left(
q_j(t)\frac{\partial^j}{\partial z^j}+\bar
q_j(t)\frac{\partial^j}{\partial \bar z^j}\right), \quad \bar q=q
\label{q}
\end{equation}
Let $\phi(z,\bar z)$ be a time-independent function satisfying
\begin{equation}
\nabla\kappa\eta\nabla\phi=0, \quad z\in\Omega \label{elliptic}
\end{equation}
in whole $\Omega$, including $z=z_1$.

Let us now estimate the time derivatives of the following quantities
$$
M[\phi]=\int_{\Omega(t)}\eta\phi dxdy.
$$
Considering an infinitesimal variation of the fluid domain
$\Omega(t)\to \Omega(t+dt)$, we get
$$
\frac{d M[\phi]}{dt}=\oint_{\partial \Omega(t)}v_n\eta\phi dl,
$$
where $dl$ is the boundary arc length. From (\ref{darcy}),
(\ref{boundary}), (\ref{kinematik}) it follows
$$
\frac{d M[\phi]}{dt}=\oint_{\partial
\Omega(t)}\left(P\kappa\eta\nabla\phi-\phi\kappa\eta\nabla
P\right)\cdot ndl.
$$
Applying the Stokes theorem and remembering that $P$ and $\phi$
satisfy (\ref{pressure}), (\ref{elliptic}), we get
\begin{equation}
\frac{dM[\phi]}{dt}=\pi \hat q(t)[\phi](z_1,\bar z_1)
\label{dynamics}.
\end{equation}
Note that mixed derivatives are absent $\hat q$ (c.f. (\ref{q})),
for by (\ref{elliptic}), $\frac{\partial^2\phi}{\partial z\partial
\bar z}$ is expressed through first derivatives of $\phi$. $\bar
q_j$ is the complex conjugate of $q_j$, since both $\phi(z,\bar z)$
and $\bar\phi(\bar z,z)$ satisfy (\ref{elliptic}).

It follows that $M[\phi]$ is conserved for any solution of
(\ref{elliptic}), a such that $\hat q[\phi](z_1,\bar z_1)=0$.

The conservation laws have been first obtained for the homogeneous
medium flows in \cite{R}, the variable-coefficient generalization
seems to be first presented in \cite{EE}.

The flow in the homogeneous medium
\begin{equation}
\kappa=1, \quad \eta=1 \label{homogeneous}
\end{equation}
is the simplest example, where the conservation laws can be written
down explicitly \cite{R}. In this example, any (anti)analytic in
$\Omega$ function satisfies (\ref{elliptic})
\begin{equation}
\phi(z, \bar z)=f(z)+g(\bar z), \quad {\rm for} \quad \kappa=1,
\quad \eta=1 \label{fu},
\end{equation}
where $f,g$ are univalent in $\Omega$ and the quantities
$$
\int_{\Omega(t)}\left(f(z)(z-z_1)^k+g(\bar z)(\bar z-\bar
z_1)^k\right)dxdy
$$
are integrals of motion for the free-boundary flows driven by a
multipole source of order $k+1$ located at $z=z_1$ in homogeneous
medium.

Returning to the general case, we integrate (\ref{dynamics}) getting
$$
M[\phi](t)=M[\phi](0)+\pi \hat Q[\phi](z_1,\bar z_1),
$$
where
\begin{equation}
\hat Q=\int_{0}^{t} \hat q(t')dt'= Q+\sum_{j=1}^k\left(
Q_j\frac{\partial^j}{\partial z^j}+\bar
Q_j\frac{\partial^j}{\partial \bar z^j}\right), \quad
\label{integrals}.
\end{equation}
Therefore $M[\phi](t)$, and consequently a form of the domain, does
not depend on the history of the sources and is a function of
``multipole fluxes"
$$
Q=\int_0^t q(t')dt', , \quad Q_j=\int_{0}^tq_j(t')dt', \quad \bar
Q_j=\int_{0}^t\bar q_j(t')dt', \quad j=1..k
$$
injected by time $t$.

Now consider the special case when $M[\phi](0)=0$ that describes the
injection of the fluid to an initially empty medium. In such a case
\begin{equation}
\int_\Omega \eta(z,\bar z)\phi(z,\bar z)dxdy=\pi \hat
Q[\phi(z_k,\bar z_k)], \label{simple}
\end{equation}
Equation (\ref{simple}) is a generalization of quadrature identities
(those expressing integrals over $\Omega$ through evaluation of
integrands and a finite number of their derivatives at a finite
number of points inside $\Omega$) appearing in the theory of
harmonic functions \cite{S} to the case of elliptic equations with
variable coefficients. Special domains for which the quadrature
identities hold are called quadrature domains in the theory of the
harmonic functions. We extend this definition to solutions of an
arbitrary elliptic PDE with regular in $\Omega$ coefficients.

To construct quadrature domains (or equivalently the domains formed
by fluid injected into initially empty porous medium), one needs an
explicit form of a general solution of (\ref{elliptic}). Such
explicit solutions are available for a class of the second order
differential equations that are related to the Schrodinger operators
of integrable systems on the plane. We, however, postpone
construction of these solution to Section \ref{calogero} and first
present our main result in the next section.

\end{section}

\begin{section}{The Main Result}\label{main}

Let us start with the simplest possible example, where the liquid is
injected in the initially empty homogeneous porous medium through
the single monopole source at $z=z_1=x_1+iy_1$. By symmetry of the
problem the solution is a circular disc of the radius $r(t)$,
centered at $z=z_1$
\begin{equation}
|z-z_1|< r(t) \label{circular}.
\end{equation}
The pressure satisfies
$$
\Delta P=-\pi {\tilde q}(t)\delta(x-x_1)\delta(y-y_1),
$$
where the source power and the total flux are
$$
{\tilde q}(t)=\frac{dr(t)^2}{dt}, \quad {\tilde Q}=r^2,
$$
respectively.

The remarkable fact is that the variable-coefficient problem, with
the constant porosity and the permeability that varies as an inverse
square of one Cartesian coordinate
\begin{equation}
\kappa=\frac{1}{x^2}, \quad \eta=1 \label{two}
\end{equation}
admits the same circular solution (\ref{circular}) if the flow is
driven by a combination of the same monopole source (of strength
$q=\tilde q(t)$) and a dipole source of strength $q_1=-{\tilde
q}{\tilde Q}/2x_1$, both located at the point $z=z_1$
\begin{equation}
\nabla\frac{1}{x^2}\nabla P=-\pi \hat q[\delta(x-x_1)\delta(y-y_1)],
\quad \hat
q=\frac{dr^2}{dt}\left(1-\frac{r^2}{2x_1}\frac{\partial}{\partial
x}\right) \label{circular2}.
\end{equation}
Indeed, it is not difficult to check that
\begin{equation}
P=r\frac{dr}{dt}\left((2x_1x+\rho^2+r^2)
\log\rho-\frac{r^2x(x-x_1)}{\rho^2}-\rho^2+x(x-x_1)-(2x_1x+\rho^2)\log(r)\right),
\label{px1}
\end{equation}
where $\rho=|z-z_1|$, satisfies (\ref{circular2}), and the boundary
conditions (\ref{darcy}), (\ref{two}),
$$
\frac{dr}{dt}=-\frac{1}{x^2}\left(\frac{\partial P}{\partial
n}\right)_{\rho=r}
$$ as well as (\ref{pressure}) holds at the disc boundary.

Also the following quadrature identity holds
$$
\int_{|z-z_1|<r}\phi dxdy=\pi r^2\phi(z_1,\bar z_1)+\frac{\pi
r^4}{4x_1}\left(\frac{\partial \phi}{\partial x}\right)_{z=z_1},
$$
which is a simplest non-trivial generalization of the mean value
theorem
 for harmonic functions to the case of regular in
(\ref{circular}) solutions of the elliptic PDE
$$
\nabla \frac{1}{x^2}\nabla \phi=0.
$$
The above example is a special case of the main result presented
below.

{\it Let $\Omega(t)$ be a domain, resulted from the injection of the
fluxes $\tilde Q,{\tilde Q}_j, j=1..{\tilde k}$ into a {\bf
homogeneous} medium through a multipole source of order ${\tilde
k}+1$ located at $z=z_1$. Then the same domain can be formed by an
injection of a special combination of fluxes $Q, Q_j, j=1..k$
through a multipole source of order $k=({\tilde k}+1)(s(n+l)+1)$
located at $z=z_1$ into an initially empty {\bf nonhomogeneous}
medium with permeability
\begin{equation}
\kappa=\frac{1}{(z^s+\bar z^s)^{2n}(z^s-\bar z^s)^{2l}}, \quad
n>l\ge 0, \quad s>0, \quad \eta=1 \label{dihedral}
\end{equation}
and constant porosity, where $s,n,l$ are integers}.

In more details, the multipole fluxes of nonhomogeneous medium
problem must be fixed functions of fluxes of its homogeneous medium
counterpart
$$
Q={\tilde Q}, \quad Q_j=Q_j({\tilde Q},{\tilde Q}_1,...,{\tilde
Q}_{\tilde k},\bar{\tilde Q}_1,...,\bar{\tilde Q}_{\tilde
k},z_1,\bar z_1), \quad j=1..({\tilde k}+1)(s(n+l)+1)-1.
$$
For instance, in the above example of circular $\tilde k=0$ solution
in a medium with permeability (\ref{two}) (that is the special case
$n=1,l=0,s=1$ of (\ref{dihedral}))
$$
\quad k=(0+1)(1(1+0)+1)-1=1, \quad Q=\tilde Q, \quad
Q_1=\frac{\tilde Q^2}{4x_1}.
$$
Note that (\ref{dihedral}) can be rewritten in the form
$$
\kappa=\frac{1}{\zeta(x,y)^2}, \quad
\zeta(x,y)=\prod_{\alpha\in{\mathcal R}_+}(\alpha\cdot
z)^{m_\alpha}, \quad (a\cdot b):={\rm Re}(\bar a b),
$$
where ${\mathcal R}=\{\alpha\}$ is a set of root vectors of a finite
reflection (Coxeter) group on the plane (Dihedral group). ${\mathcal
R}$ is invariant under reflections $z\to z-2\frac{(\alpha\cdot
z)}{(\alpha\cdot\alpha)}\alpha$ in a mirror (line), normal to any
root vector $\alpha \in {\mathcal R}$, and ${\mathcal R}_+$ denotes
a positive subset of ${\mathcal R}$, containing a half of all root
vectors. The multiplicities $m_\alpha$ must be non-negative integers
that are functions on the group orbits. If $l=0$, the reflection
group has one orbit and $m_\alpha=n$. Otherwise, the group has two
orbits and multiplicities $m_\alpha$ take values $n$ and $l$ on each
orbit respectively. The permeability (\ref{dihedral}) is therefore
an invariant of the group of symmetries of a regular $4s$-polygon
($2s$-polygon if $l=0$) and its singular locus coincides with the
union of mirrors.

The pressure satisfies the elliptic PDE
$$
\nabla\zeta(x,y)^{-2}\nabla P=\zeta(x,y)^{-1}H\zeta(x,y)^{-1}P=-\pi
\hat q[\delta(x-x_k)\delta(y-y_k)]
$$
$$
{\rm order}(\hat q)=({\tilde
k}+1)\left(\deg(\zeta(x,y))+1\right)-1=({\tilde
k}+1)\left(1+\sum_{\alpha\in{\mathcal R}_+}m_\alpha\right)-1,
$$
where $H$ is the Schr\"{o}dinger operator of the Calogero-Moser
system related to the dihedral group \cite{OP}, \cite{CFV},
\cite{BL}
$$
H=\Delta-\sum_{\alpha\in{\mathcal
R}_+}(\alpha\cdot\alpha)\frac{m_\alpha(m_\alpha+1)}{(\alpha\cdot
z)^2}.
$$
Also, regular in $\Omega$ solutions $\phi(z,\bar z)$ of the elliptic
PDEs $\nabla\zeta(x,y)^{-2}\nabla \phi=0$ satisfy the quadrature
identities
$$
\int_{\Omega}\phi dx dy=\hat Q[\phi](z_1,\bar z_1), \quad {\rm
order}(\hat Q)=({\tilde k}+1)\left(\deg(\zeta(x,y))+1\right)-1
$$
that are generalization of the mean value theorem for harmonic
functions to the case of solutions of the variable-coefficient
elliptic PDEs in (generally) non-circular domains.

Before proving the main result we need to construct a complete set
of solutions to (\ref{elliptic}) for the medium with $\kappa, \eta$
given by (\ref{dihedral}).

\end{section}

\begin{section}{Nonhomogeneous porous medium flows and integrable systems related to the finite reflection
groups}\label{calogero}

In this section we show how to obtain a general solution
$\phi(z,\bar z)$ to (\ref{elliptic}), (\ref{dihedral}). It is
instructive to consider the special case $s=1, l=0$ of
(\ref{dihedral})
\begin{equation}
\kappa=\frac{1}{x^{2n}}, \quad \eta=1 \label{one}
\end{equation}
when permeability depends on one Cartesian coordinate only.
Construction of solutions in the general case (\ref{dihedral}) is
conceptually similar.

Consider the simplest non-trivial example $n=1$ in (\ref{one}) and
start with factorizing the differential operator $\partial_x^2$ as
$$
\frac{\partial^2}{\partial
x^2}=\left(\frac{1}{x}\frac{\partial}{\partial
x}\right)\left(x\frac{\partial}{\partial x}-1\right).
$$
By the associativity of the differential operators
$$
\begin{array}{ccccc}
\left(x\frac{\partial}{\partial x}-1\right)
&\underbrace{\left(\frac{1}{x}\frac{\partial}{\partial
x}\right)\left(x\frac{\partial}{\partial x}-1\right)}& = &
\underbrace{\left(x\frac{\partial}{\partial
x}-1\right)\left(\frac{1}{x}\frac{\partial}{\partial x}\right)} &
\left(x\frac{\partial}{\partial
x}-1\right)\\
 &\partial_x^2 & & x^2\partial_x\frac{1}{x^2}\partial_x &.
\end{array}
$$
Therefore,
\begin{equation}
T_1\Delta=L_1T_1, \quad T_1=x\frac{\partial}{\partial x}-1, \quad
L_1=x^2\nabla\frac{1}{x^2}\nabla \label{example}
\end{equation}
and the elliptic equation (\ref{elliptic}) is nontrivially related
to the Laplace equation when $\kappa\eta=1/x^2$. In general, the
identity
$$
T\Delta=LT
$$
that relates two differential operators (e.g. $\Delta$ and $L$
above) through a differential operator $T$ is called an intertwining
identity and $T$ is an intertwining operator.

A differential operator that is related to $\Delta$ through the
intertwining identity equals, modulo a gauge transformation, a
Schr\"{o}dinger operators of an integrable system. Indeed,
$$
T\left(\Delta-\lambda\right)=\left(L-\lambda\right)T
$$
and when $\nu$ is an eigenfunction of $\Delta$ with the eigenvalue
$\lambda$, $T[\nu]$ is an eigenfunction of $L$ with the same
eigenvalue or zero.

The factorization approach leading to the simplest nontrivial
intertwining identity (\ref{example}) can be now applied to $L_1$
etc. By induction we get the intertwining identity for an arbitrary
nonnegative integer $n$ in (\ref{one})
\begin{equation}
T_n\Delta=L_nT_n, \quad L_n=x^{2n}\nabla\frac{1}{x^{2n}}\nabla
\label{intertwining1},
\end{equation}
where
\begin{equation}
T_n=x^n\left(\frac{\partial}{\partial
x}-\frac{n}{x}\right)\left(\frac{\partial}{\partial
x}-\frac{n-1}{x}\right)...\left(\frac{\partial}{\partial
x}-\frac{1}{x}\right)=\sum_{i=0}^na_{i;n}x^i\frac{\partial^i}{\partial
x^i} \label{Tn}.
\end{equation}
Any solution $\phi$ to (\ref{elliptic}), (\ref{one}) in $\Omega$ can
be represented in the form
\begin{equation}
\phi=T_n[f], \quad \Delta f=0, \quad z\in \Omega \label{proof}.
\end{equation}
Let us show this for $n=1$, where $T_1$ is given by (\ref{example}).
Introduce $f$ satisfying the following equation
$$
x\frac{\partial f}{\partial x}-f=\phi,
$$
where $L_1[\phi]=0$ in $\Omega$ and $\phi=0$ for $z\not\in\Omega$.
It is not difficult to see that
$$
f(z,\bar
z)=x\int_{-\infty}^x\frac{\phi(x'+iy,x'-iy)}{(x')^2}dx'+xF(y),
$$
where $F(y)$ is an arbitrary regular function of $y$. Therefore, for
any $\phi$ regular in $\Omega$ (which is the case) there exist a
regular in $\Omega$ function $f$, such that $T_1[f]=\phi$. From the
intertwining identity (\ref{example}) it follows that $ T_1\Delta
f=0 $ and $\Delta f \in {\rm Ker}(T_1)$, if $z\in \Omega$ i.e.
$$
\Delta[f]=xC(y), \quad z=x+iy \in \Omega
$$
where $C(y)$ is an arbitrary function of $y$. It follows that
$$
\Delta[f+xF(y)]=0, \quad \frac{d^2F(y)}{dy^2}=C(y).
$$
Since $f$ is defined modulo $xF(y)$, we can set $F=0$ and any
solution of $L_1[\phi]=0$ can be represented as $\phi=T_1[f]$, where
$\Delta[f]=0$. Similar proof applies to the arbitrary $n$ case
(\ref{proof}).

In the general case (\ref{dihedral})
$$
T_{n,l;s}\Delta=L_{n,l;s}T_{n,l;s}, \quad L_{n,l;s}=(z^s+\bar
z^s)^{2n}(z^s-\bar z^s)^{2l}\nabla \frac{1}{(z^s+\bar
z^s)^{2n}(z^s-\bar z^s)^{2l}}\nabla,
$$
where the intertwining operator can be expressed in the form of a
Wronskian \cite{BL}
\begin{equation}
T_{n,l;s}[f]=\rho^{s(n+l)}\frac{W[\sin(\theta_1),\sin(\theta_2),...,\sin(\theta_n),f]}{\cos(s\theta)^{n(n-1)/2}\sin(s\theta)^{l(l-1)/2}},\quad
W[f_1,..,f_k]:=\det\left[\frac{\partial^{j-1} f_i}{\partial
\theta^{j-1}}\right]_{1\le i,j\le k} \label{intertwining}
\end{equation}
with
$$
z=\rho e^{i\theta}, \quad \theta_k=\left\{\begin{array}{cc}
k\left(s\theta+\frac{\pi}{2}\right),& k=1,2,..n-l \\
(2k+l-n)\left(s\theta+\frac{\pi}{2}\right),& n-l<k\le n
\end{array}.
\right.
$$
It is important that the intertwining operator $T_{n,l;s}$ is a
homogeneous differential polynomial in $z,\bar z,
\partial_z, \partial_{\bar z}$. This fact allows one to construct the quadrature domains for solutions $\phi$ of $L_{n,l;s}[\phi]=0$.

Note that in general, there exist several independent operators
intertwining $\Delta$ and another second order differential operator
(forming a linear space of intertwining operators). For instance,
$T_n$ in (\ref{intertwining1}) that intertwines $\Delta$ with
$L_n=L_{n,0;1}$ is not a special case $T_{n,0;1}$ of
(\ref{intertwining}). However, any nonzero linear combination of
them can be used to obtain solutions to the corresponding elliptic
equations $L_{n,0;1}[\phi]=0$.

\end{section}

\begin{section}{Proof of the main result}\label{final}

The zero-initial condition solution of the homogeneous-porous medium
flow that is driven by a multipole source of order ${\tilde k}+1$ is
described by the polynomial conformal map of degree ${\tilde k}+1$
from the unit disc in the parametric $|w|$-plane into the fluid
region $\Omega$
\begin{equation}
z(w)=z_1+rw+\sum_{i=1}^{\tilde k}u_iw^{i+1}, \quad |w|<1
\label{polynomial}
\end{equation}
The map is analytic in $|w|<1$, and the unit circle $|w|=1$ is
mapped to the boundary $\partial\Omega$. As shown above such regions
are also quadrature domains. They are special (polynomial) cases of
(rational) algebraic domains \cite{VE}. The map coefficients $r,
u_i,i=1,..{\tilde k}$ are functions of $\tilde Q, \tilde Q_i, \bar
{\tilde Q}_i, i=1..{\tilde k}$. The ``conformal radius" $r$ can be
chosen to be real.

The main idea of the proof is to show that the quadrature identity
(\ref{simple}) for solutions of (\ref{elliptic}), (\ref{dihedral})
holds in domains defined by (\ref{polynomial}).

Consider illustrative examples of the problem (\ref{one}) with
permeability changing in one direction. As shown in the previous
Section any solution of the elliptic equation $ L_n[\phi]=0$ can be
represented as
\begin{equation}
T_n[f(z)+g(\bar z)], \label{tfg}
\end{equation}
where $f(z), g(\bar z)$ are holomorphic and anti-holomorphic
respectively.

According to (\ref{simple}) we have to show that
$$
\int_\Omega T_n[f(z)]dxdy=\pi\hat QT_n[f(z)]_{z=z_1}
$$
holds for any analytic in $\Omega$ function $f(z)$ when $\Omega$ is
defined by the conformal map (\ref{polynomial}). Using the Green
theorem and taking (\ref{Tn}) into account we rewrite the last
equation as
\begin{equation}
\sum_{j=0}^n\frac{a_{j;n}}{2^j(j+1)}\frac{1}{2\pi
i}\oint_{\partial\Omega}(z+\bar
z)^{j+1}\frac{\partial^jf(z)}{\partial z^j}dz=\hat
QT_n[f(z)]_{z=z_1} \label{algebraic}.
\end{equation}
Since $\bar w=1/w$ if $|w|=1$, $\bar z(\bar w)=\bar z(1/w)$ along
the boundary, and we can rewrite the left-hand side of the last
equation as
$$
\sum_{j=0}^n\frac{a_{j;n}}{2^j(j+1)}\frac{1}{2\pi
i}\oint_{|w|=1}\left(\left(z(w)+\bar
z(1/w)\right)^{j+1}\left(\frac{1}{\frac{\partial z(w)}{\partial
w}}\frac{\partial }{\partial
w}\right)^j[f(z(w))]\right)\frac{\partial z(w)}{\partial w}dw.
$$
Since $z(w)$ is analytic in $|w|<1$, $\bar z(1/w)$ has poles only at
$w=0$, and the above integral is a pure sum of residues
\begin{equation}
\sum_{j=0}^{({\tilde
k}+2)(n+1)-2}V_j\left(\frac{\partial^jf(z)}{\partial
z^j}\right)_{z=z_1}, \label{lhs}
\end{equation}
where $V_j, j=0..({\tilde k}+2)(n+1)-2$ are functions of the
parameters $z_1, \bar z_1, r,u_j,\bar u_j, j=1..\tilde k$ of the
conformal map (\ref{polynomial}). Equating it with the right-hand
side of (\ref{algebraic}), we see that $\hat Q_k$ must be
(differential operators) of order $({\tilde k}+1)(n+1)-1$ and
\begin{equation}
\hat QT_n[f(z)]_{z=z_1}=\sum_{j=0}^{({\tilde
k}+2)(n+1)-2}U_j\left(\frac{\partial^jf(z)}{\partial
z^j}\right)_{z=z_1}, \label{rhs}
\end{equation}
where $U_j, j=0..({\tilde k}+2)(n+1)-2$ are linear functions of
$Q,Q_j, \bar Q_j, j=1..({\tilde k}+1)(n+1)-1$. Therefore, the
quadrature identity (\ref{algebraic}) is satisfied if the following
system of $2({\tilde k}+2)(n+1)-1$ linear equations
\begin{equation}
V_j-U_j=0, \quad \bar V_j-\bar U_j=0, \quad j=0..({\tilde
k}+2)(n+1)-2 \label{defining}
\end{equation}
for $2({\tilde k}+1)(n+1)$ unknowns
\begin{equation}
Q, \bar Q, Q_j,\bar Q_j, \quad j=1..({\tilde k}+1)(n+1)-1 \label{Q}
\end{equation}
has solutions.

Note that the condition $\bar Q=Q$ is satisfied automatically,
since, as easily seen from (\ref{algebraic}), (\ref{polynomial}),
(\ref{integrals}), the $j=0$ subset of (\ref{defining})
$$
V_0-U_0=0, \quad \bar V_0-\bar U_0=0
$$
are equations for $Q, \bar Q$ that have the same form for any $n\ge
0$ in (\ref{Tn}). They have real solution
$$
Q=\tilde Q=r^2+\sum_{i=1}^{\tilde k}(i+1)u_i\bar u_i
$$
The number of equations in (\ref{defining}) exceeds the number of
unknowns (\ref{Q}) by $2n$ and the system of equations
(\ref{defining}) is overdermined for the non-homogeneous medium
problem $n>0$.

For instance, for the circular domain $\tilde k=0$
$$
z(w)=z_1+rw
$$
in a medium with permeability $1/x^2$ (that has been considered in
Section \ref{main}), system (\ref{defining}) consists of six
equations
$$
\begin{array}{l}
Q-r^2=0, \quad \bar Q-r^2=0, \\
2Qx_1-Q_1+\bar Q_1-2x_1r^2=0, \quad 2\bar Qx_1-\bar Q_1+Q_1-2x_1r^2=0, \\
4Q_1x_1-r^4=0, \quad 4\bar Q_1x_1-r^4=0
\end{array}
$$
for four unknowns $Q,\bar Q, Q_1, \bar Q_1$. It has the following
solution (already demonstrated in Section \ref{main})
$$
Q=\bar Q=r^2, \quad Q_1=\bar Q_1=r^4/4x_1
$$
Returning to the general case, we are going to show that not all
equations in (\ref{defining}) are independent, the system is
compatible and has a unique solution, which proves our main result
presented in Section \ref{main}.

To prove the compatibility, we introduce the basis
\begin{equation}
\phi_j(z,\bar z)=c_jT_n[(z-z_1)^{j+n}], \quad \bar \phi_j(\bar z,
z)=c_jT_n[(\bar z-\bar z_1)^{j+n}], \quad j=0,1,2,... \label{basis}
\end{equation}
where $c_j=j!/(j+n)!x_1^n$, in the space of solutions $\phi(z,\bar
z)$ of $L_n[\phi]=0$ that are regular in neighborhood of $z=z_1$,
and then show that the quadrature identity (\ref{simple}) holds for
any element of this basis.

Indeed, according to (\ref{Tn}), (\ref{basis}) is a set of solutions
of $L_n[\phi]=0$ that continuously tends to the basis of functions
analytic in a neighborhood of $z=z_1$
$$
\phi_j(z,\bar z)\to (z-z_1)^j, \quad j=0,1,2,..., \quad x_1\to
\infty
$$
as the position of the source $z_1=x_1+iy_1$ goes to infinity. On
the other hand, $L_n[\phi(z,\bar z)]=0, z \in \Omega$ continuously
tends to the Laplace equation when region $\Omega$ is moved to
infinity. Therefore, set (\ref{basis}) is homotopically equivalent
to $(z-z_1)^j, (\bar z-\bar z_1)^j, j=0,1,2,..$ under a continuous
deformation of the Laplace operator and thus contains a basis of
solutions of $L_n[\phi]=0$ that are regular in a neighborhood of
$z=z_1$.

It then follows from (\ref{lhs}) and (\ref{rhs}) that the quadrature
identity holds if
\begin{equation}
(V_j-U_j)\left(\frac{d^jf(z)}{dz^j}\right)_{z=z_1}=0, \quad (\bar
V_j-\bar U_j)\left(\frac{d^jf(\bar z)}{d\bar z^j}\right)_{z=z_1}=0,
\quad j=0..({\tilde k}+2)(n+1)-2 \label{uv}
\end{equation}
for
$$
f(z) \in \{(z-z_1)^j, j\ge n\}
$$
Since $j\ge n$, the left hand sides of the first $n$ equations in
(\ref{uv}) vanish, and there remains $2({\tilde k}+1)(n+1)$
independent equations and the equal number of unknowns (\ref{Q}).

We now have to prove the compatibility of remaining equations.
(\ref{defining}) is a non-homogenous system of linear equations for
unknowns $Q, \bar Q, Q_j,\bar Q_j, j=1..({\tilde k}+1)(n+1)-1$, that
fixes dependence of these unknowns on parameters $z_1,\bar
z_1,r,u_1,...,u_{\tilde k},$ $\bar u_1,...,\bar u_{\tilde k}$. The
system is compatible if its homogenous part does not have nontrivial
solutions. Let us suppose that it does. Remind that the homogenous
part of the system has been obtained by action of the operator $\hat
Q$ to an arbitrary solution of $L_n[\phi]=0$ at point $z=z_1$. So,
if the homogeneous part of the system had nontrivial solutions, then
operator $\hat Q$ would annihilate any solution $\phi$ of
$L_n[\phi]=0$ at $z=z_1$, i.e.
\begin{equation}
\hat Q T_n[f(z)]=0 \quad {\rm at} \quad z=z_1, \label{action}
\end{equation}
where $f(z)$ is any analytic in $\Omega$ function. If the above were
true, then changing continuously the position $z=z_1$ of the source,
we could construct such operator $\hat Q$, with coefficients
depending on $z$, that $\hat QT_n[f(z)]=0$ in some region of the
plane for an arbitrary $f(z)$. But this is evidently impossible,
since the highest symbols of $\hat Q$ and $T_n$ contain pure
derivatives in $z$, so does their composition.

Therefore (\ref{defining}) has a unique solution and quadrature
identity holds in the polynomial algebraic domains, that leads to
our main result (see Section 3).

Similar result can be analogously proved for the general system with
permeability given by (\ref{dihedral}).

\end{section}

\begin{section}{Generalization to Higher Dimensions}

Our study can be extended to PDEs in more than two dimensions, since
the derivation of the conservation laws in Section 2 is not
restricted to the two-dimensional flows.

For instance, if we take solution $\phi(\xi)$ of $\nabla
\xi_1^{-2}\nabla \phi$=0, where $\zeta:=(\xi_1,..,\xi_d)$ and
$\nabla:=\left(\frac{\partial}{\partial
\xi_1},..,\frac{\partial}{\partial \xi_d}\right)$ , then the
following quadrature identity holds
$$
\int_{|\xi-\xi'|<r}\phi(\xi) d
\xi_1..d\xi_d=v_d\left(\phi(\xi)+\frac{r^2}{(d+2)\xi_1}\frac{\partial
\phi(\xi)}{\partial \xi_1}\right)_{\xi=\xi'}
$$
in the $d$-dimensional ball of radius $r$ with center at $\xi=\xi'$,
where $v_d=\int_{|\xi-\xi'|<r} d \xi_1..d\xi_d$ is volume of the
ball. Since the Hadamard expansion of the fundamental solution of
the Calogero-Moser type operators truncates (see e.g. \cite{BM},
\cite{CFV}), the pressure distribution in the corresponding
free-boundary flow can be written down explicitly for any $d$ (e.g.
we have used technique of \cite{BM} for derivation of (\ref{px1})).

\end{section}

\begin{section}{Conclusion, Main Conjecture}

In this article we have found examples of non-homogeneous porous
medium flows, driven by a multipole source at a fixed point, whose
boundaries obey the same dynamics as those of the homogeneous-medium
flows also driven by a multipole source located at the same point.
Namely, the medium permeability is a homogeneous rational function
of $x,y$, and an invariant of the dihedral group. The multipole
fluxes of the non-homogeneous medium problem must be fixed functions
of those of the homogeneous medium one. The related
variable-coefficient elliptic PDEs for the pressure distribution are
of the integer multiplicity Calogero-Moser type and the quadrature
identities for solutions of such equations hold in polynomial
algebraic domains.

The above result has been derived using technique of intertwining
operators. The fact that coefficients of these operators are
polynomials in $z, \bar z$ is essential for algebraicity of the
corresponding quadrature domains. This result can be extended in the
following way

According to works on algebraic integrability \cite{CFV}, it is
likely that the second-order elliptic operators related to the
Coxeter root systems as well as their special deformations exhaust
all possible operators that can be related to the Laplace operators
through polynomial intertwining operators. These are elliptic
operators for which a polynomial Baker-Akhieser function exists. We
call all such operators the Algebraic Calogero-Moser operators. For
instance, the algebraic Calogero-Moser equations \cite{CFV} related
to the porous medium problems (\ref{elliptic}) with the permeability
$$
\kappa=x^{-2m}\left((2m+1)y^2-x^2\right)^{-2}, \quad \eta=1,
$$
where $m=2,3,4,...$, are simplest examples that extend the main
result of the present paper to non-Coxeter arrangements of mirrors.

More generally, the problem of classification of all such algebraic
Calogero-Moser systems in two dimensions is as follows \cite{B},
\cite{BL}, \cite{CFV}: Find a strictly increasing sequence of
integer positive numbers $0\le k_1<k_2<..<k_n$ and a sequence
$\omega_1,..\omega_n$ of complex parameters (``phases"), such that
the ratio of Wronskians
$$
\zeta(x,y)=\rho^{k_n}\frac{W[\sin(\theta_1),..\sin(\theta_{n-1}),\sin(\theta_n)]}
{W[\sin(\theta_1),..\sin(\theta_{n-1})]}, \quad
\theta_j=k_j\theta+\omega_j, \quad z=\rho e^{i\theta}
$$
is a polynomial in $x,y$. The corresponding intertwining operator is
of the $n$th order and a polynomial in $x,y$ that can be rewritten
in the form of a Wronskian
$$
T[f]=\rho^{k_n}\frac{W[\sin(\theta_1),...\sin(\theta_n),f]}{W[\sin(\theta_1),...\sin(\theta_{n-1})]}
$$
with (\ref{intertwining}) being a special case. The results of the
present paper (i.e. algebraicity of domains for PDEs with
$\kappa=1/\zeta(x,y)^2$) also hold for the deformed systems.

In conclusion, one may pose the classification problem: {\it Find
complete list of all PDEs whose solutions satisfy quadrature
identities in polynomial algebraic domains}. In view of the above it
is reasonable to expect that it has the following solution:

{\bf Conjecture}: {\it The algebraic Calogero-Moser equations
exhaust, up to a gauge equivalence, all possible second order
elliptic PDEs whose solutions satisfy quadrature identities in
polynomial algebraic domains}

As mentioned in Section 6, our study can be extended to PDEs in more
than two dimensions. Although, we cannot use conformal maps to
parametrize algebraic domains in higher dimensions, we can still
pose the classification problem for domains with spherical
boundaries, looking for a complete list of elliptic equations for
which generalized mean value theorem holds. More precisely, one can
look for equations in $d$-dimensions whose solutions possess the
following property: {\it Integral of an arbitrary solution taken
over an arbitrary $d$-dimensional ball, equals a linear combination
(with coefficients depending on the ball radius and position) of the
value of the solution and those of a finite number of its
derivatives at the ball center}.

Extending our conjecture to higher dimensions we may expect that
algebraic Calogero-Moser equations complete the above list. Note,
that our classification problem in $d>2$ seems to be equivalent to a
restricted classical Hadamard problem for irreducible Huygens'
operators that non-trivially depend on more than two variables. As
in our case, all known examples of such Huygens' operators are
related to algebraic Calogero-Moser systems \cite{CFV}.

\end{section}

\begin{section}{Acknowledgements}

The authors are grateful to Y. Berest, P.Etingof, J. Harnad, S.
Howison, and J. Ockendon for stimulating discussions and useful
remarks. The work of O.Y. and I.L. was supported by the European
Community IIF's MIF1-CT-2004-007623 and MIF1-CT-2005-007323.

\end{section}

\end{document}